# How Kondo Holes Create Intense Nanoscale Heavy-Fermion Hybridization Disorder

M.H. Hamidian[a,b,§], A.R. Schmidt[a,b,c,§], I.A. Firmo[a,b], M.P. Allan[a,d], P. Bradley[e], J.D Garrett[f], T.J. Williams[g], G.M. Luke[g,h], Y. Dubi[i,j], A.V. Balatsky[i], and J.C. Davis[a,b,d,k]

[a]Laboratory for Atomic and Solid State Physics, Department of Physics, Cornell University, Ithaca, NY 14853, USA; [b]Condensed Matter Physics and Materials Science Department, Brookhaven National Laboratory, Upton, NY 11973, USA; [c]Department of Physics, University of California, Berkeley, CA 94720, USA; [d]Scottish Universities Physics Alliance (SUPA), School of Physics and Astronomy, University of St Andrews, North Haugh, St Andrews KY16 9SS, UK; [e]Department of Physics, University College Cork, Co Cork, Ireland; [f]Brockhouse Institute for Materials Research, McMaster University, Hamilton, ON, Canada, L8S 4M1; [g]Department of Physics & Astronomy, McMaster University, Hamilton, Ontario, L8S 4M1 Canada; [h]Canadian Institute for Advanced Research, Toronto, Ontario, M5G 1Z8, Canada; [i]Theory Division, Los Alamos National Lab., Los Alamos, NM 87545, USA; [j]Sackler School of Physics and Astronomy, Tel Aviv University, Tel Aviv 69978, Israel; [k]Kavli Institute at Cornell for Nanoscale Science, Cornell University, Ithaca, NY 14850, USA.

[§]These authors contributed equally to this project.

**Replacing a magnetic atom by a spinless atom in a heavy fermion compound generates a quantum state often referred to as a 'Kondo-hole'. No experimental imaging has been achieved of the atomic-scale electronic structure of a Kondo-hole, or of their destructive impact** (Lawrence JM, et al. (1996) Kondo hole behavior in $Ce_{0.97}La_{0.03}Pd_3$. *Phys Rev B* 53:12559-12562; Bauer ED, et al. (2011) Electronic inhomogeneity in a Kondo lattice. *Proc Natl Acad Sci.* 108:6857-6861) **on the hybridization process between conduction and localized electrons which generates the heavy fermion state. Here we report visualization of the electronic structure at Kondo-holes created by substituting spinless Thorium atoms for magnetic Uranium atoms in the heavy-fermion system $URu_2Si_2$. At each Thorium atom, an electronic bound state is observed. Moreover, surrounding each Thorium atom we find the unusual modulations of hybridization strength recently predicted to occur at Kondo-holes** (Figgins J, Morr DK (2011) Defects in heavy-fermion materials: unveiling strong correlations in real space. *Phys Rev Lett* 107:066401). **Then, by introducing the 'hybridization gapmap' technique to heavy fermion studies, we discover intense nanoscale heterogeneity of hybridization due to a combination of the randomness of Kondo-hole sites and the long-range nature of the hybridization oscillations. These observations provide direct insight into both the microscopic processes of heavy-fermion forming hybridization and the macroscopic effects of Kondo-hole doping.**

**Kondo Lattice generated Heavy Fermions**. Within a Kondo lattice (Fig. 1*A*), an *f*-electron state localized at each magnetic atom becomes hybridized at low temperatures with the delocalized conduction electrons of the metal (1-4). The resulting 'heavy fermion' electronic structure in momentum space (**k**-space) is shown schematically in Fig. 1*B*. Here the original light conduction band $\varepsilon^c(\mathbf{k})$ splits into two distinct heavy bands within an energy range often referred to as the direct (4) 'hybridization gap' $\Delta_h$ (Fig. 1*B,C*). This energy range is determined by the strength of the matrix element $V_h(\approx \Delta_h/2)$ for the quantum conversion of a conduction electron into an *f*-electron and vice versa, and (in simplest models) is not centered on the Fermi energy but on the renormalized energy level of the localized *f*-electron $\varepsilon^f$ (Fig. 1*B,C* and refs 2-4). The value of $\Delta_h$ is defined quantitatively (2-6) in **k**-space as the energy-separation between the heavy fermion bands at the wavevector $\mathbf{k}^c(\varepsilon^f)$ where $\varepsilon^f$ intersects the light conduction band (Fig. 1*B*), and equivalently in real-space (**r**-space) as the energy range within which changes occur in the density of states due to appearance of the heavy fermion bands (Fig. 1*C*). We note that $\Delta_h$ is not the energy gap of a symmetry-breaking electronic phase transition, but merely the energy range of a quantum mechanical avoided crossing in **k**-space (Fig. 1*B,C*). Very recently it has become possible to detect both the characteristic **k**-space (7) and **r**-space (8) signatures of this heavy-fermion hybridization gap by using spectroscopic imaging scanning tunneling microscopy (SI-STM), with the observations being in excellent agreement with theoretical expectations.

**Kondo Resonances and Kondo Holes.** Heavy-fermions owe their existence to a special quantum mechanical many-body state that appears at a magnetic rare-earth or actinide atom when embedded in a metal. The atom's magnetic moment is screened by the formation of a spin-zero many-body electronic state (1-4), often referred to as the 'Kondo resonance' (blue atom Fig. 2*A*). Kondo resonances have been imaged directly at individual magnetic adatoms on metal surfaces (9-11) by using SI-STM to map the differential tunneling conductance $dI/dV(\mathbf{r}, \varepsilon = ev) \equiv g(\mathbf{r}, \varepsilon) \propto N(\mathbf{r}, \varepsilon)$. Here *v* is the tip-sample bias voltage for electron tunneling and $N(\mathbf{r}, \varepsilon)$ is the electronic density-of-states. A distinct type of many-body state, which can be considered dual to the Kondo resonance, occurs when a single non-magnetic atom is substituted within the array of otherwise magnetic atoms of a heavy fermion system (green atom Fig. 2*B*). This state is often referred to as a 'Kondo-hole'. Although experimentally there is good understanding of the electronic structure at individual Kondo resonances (9-11), the atomic-scale electronic phenomena surrounding Kondo-holes have remained unexplored.

**Electronic Structure of a Kondo Hole**. Understanding the electronic structure of a Kondo-hole is of fundamental and practical consequence to the study of rare-earth and actinide heavy fermion metals and superconductors. The classic experimental approach is to determine the macroscopic response to random Kondo-hole doping (12-18). A wide variety of key issues in heavy fermion physics have been studied this way including: (i) dilution and the resulting nanoscale spatial heterogeneity in the hybridization process (12-14,18); (ii) the local perturbations to the hybridization and heavy-fermion formation at individual Kondo-holes (14,17,18); (iii) local impurity states bound within the hybridization gap at the Kondo-hole, and their impact on scattering (12-15,17); (iv) the temperature minimum in resistivity possibly due to the special scattering properties of states bound within the doped Kondo-holes (14-17); (v) the impact of Kondo-holes on the development of coherence in the heavy fermion band (12,14,15,17,18); (vi) how Kondo-holes destroy unconventional superconductivity due to their special pair-breaking scattering (12-14,16-18). While the macroscopic effects of non-magnetic dopant atoms are evident throughout these studies, knowledge of the microscopic electronic phenomenology of individual Kondo-holes, and precisely how they disorder the hybridization process producing the heavy fermions, could greatly improve the precision and significance of the resulting conclusions.

**Kondo Hole Theory**. Theoretical predictions for the impact of Kondo-holes on the surrounding electronic structure and on the hybridization process generating the heavy fermions include that: (i) increasing densities of Kondo-holes will progressively degrade hybridization and its related effects on $N(\varepsilon)$ in a heterogeneous fashion at nanoscale (19-21) - eventually destroying the heavy fermion state; (ii) a special bound impurity-state should exist within the hybridization gap at each Kondo-hole site (20-22); (iii) the electrical resistivity should exhibit a minimum and then increase with falling T because of the unique scattering properties of these bound states (19-21); (iv) the heavy fermion superconductivity in non s-wave systems should be strongly suppressed by pair-breaking scattering at each Kondo-hole (23). More recent modelling of electronic structure surrounding a single Kondo-hole analyzes the atomic-scale variations in hybridization strength $V(\mathbf{r})$, the density of states $N(\mathbf{r},\varepsilon)$, and the electronic impurity-state bound inside the altered hybridization gap (22). Figure 2C shows the prediction for modulations in the strength of hybridization surrounding a Kondo-hole (22) with the key observation being that $V_h(\propto \Delta_h(\mathbf{r}))$ oscillates with wavevector $\mathbf{Q}^* \cong 2\mathbf{k}_F^c$, twice the Fermi wavevector of the original light (un-hybridized) conduction band (Fig. 2D). Such hybridization oscillations would be quite distinct from the heavy fermion Friedel oscillations in $N(\mathbf{r}, \varepsilon = 0)$ which have been observed at $\mathbf{q} = 2\mathbf{k}_F^{HF}$ (7), where $\mathbf{k}_F^{HF}$ is the Fermi wavevector of the heavy fermion band (Fig. 1B and ref. 7). An intuitive picture of this situation is that the atomic-scale hybridization suppression at the Kondo-hole generates oscillations in $\Delta_h(\mathbf{r}) \propto \delta V_h(\mathbf{r})$ at twice the characteristic wavevector $\mathbf{k}^c(\varepsilon^f)$ of the hybridization process itself (the avoided crossing of the localized f-electron state $\varepsilon^f$ and the light pre-hybridization band $\mathbf{k}^c(\varepsilon)$ - see arrow in Fig 1B). Because $\varepsilon^f$ is so close to Fermi level, $\varepsilon_F$, the resulting hybridization oscillation wavevector would be virtually indistinguishable from $2\mathbf{k}_F^c$ and therefore in good agreement with ref. 22. However, no experimental tests of any of these predictions for the electronic structure surrounding a Kondo-hole have been reported.

**Spectroscopic Imaging STM.** The long term experimental challenges have therefore been: (i) to directly visualize the electronic structure of individual Kondo-holes; (ii) to image any nanoscale heterogeneity of hybridization generated by random Kondo-hole doping; (iii) to determine the alterations to the hybridization strength and thus the heavy fermion states nearby a Kondo-hole; (iv) to search for the states bound within the hybridization gap at each Kondo-hole; (v) to explore the relationship between all these phenomena and the macroscopic transport and thermodynamic observables. In addressing these issues we use SI-STM because it can determine simultaneously the **r**-space and **k**-space electronic structure of correlated electron systems (7-11).

**Kondo Hole Doped Heavy Fermion System**. A canonical example of a Kondo-hole (14) occurs at a spinless Th atom substituted for a magnetic U atom in $URu_2Si_2$. Despite the incompletely understood 'hidden order' phase transition at ~18K in this material, the observed evolution in **k**-space upon cooling to low temperatures (~2K) consists of the rapid development of a relatively uncomplicated heavy fermion system (7,24,25). That this is a heavy fermion metal is well known from specific heat measurements giving m*~25$m_e$ (26) with little change due to thorium dilution (14), as well as from carrier damping in Hall transport data (27), and from quantum oscillations yielding m*~30$m_e$ (28,29). More directly, angle resolved photoemission studies reveal the expected signature of the **k**-space heavy fermions - the very slowly dispersing quasiparticle band of f character (24,25). Even more definitively, heavy quasiparticle interference imaging has visualized directly the splitting of the light conduction band into the two heavy bands and the associated opening of a heavy-fermion hybridization gap (7). These are all as predicted for the appearance of heavy-fermion coherence in a Kondo lattice system (2-4,30). Collectively, these observations are compelling as to the heavy fermion character of the **k**-space electronic structure in URu$_2$Si$_2$ at low temperatures. Thus, for the purposes of examining the local electronic structure of a Kondo-hole at low temperatures, URu$_2$Si$_2$ is effectively a simple heavy fermion system.

## Results

**SI-STM Measurements.** We use 1% Th-doped URu$_2$Si$_2$ samples which are well known to exhibit heavy fermion phenomena highly similar to those of the non-thorium doped compounds (14). The samples are inserted into the dilution refrigerator based SI-STM system, mechanically cleaved in cryogenic ultra-high vacuum, and then inserted into the STM head. Atomically flat and clean a-b surfaces consisting of layers of U atoms are achieved throughout. Sub-kelvin ultra high vacuum conditions enable long term $g(\mathbf{r}, \varepsilon)$ measurements with atomic resolution and register in the same field of view (FOV) without any degradation of surface quality. These $g(\mathbf{r}, \varepsilon)$ data were acquired with a standard AC lock-in amplifier technique using bias modulations down to 250 μV rms at the lowest temperatures. For **k**-space determination, a field-of-view of up to 60 nm x 60nm is used to ensure sufficiently high Fourier resolution in the dispersive scattering interference **q**-vectors (7).

**Kondo Holes and Heavy Fermion Hybridization.** In Fig. 3*A* we show a topographic image of the Uranium-termination surface (7) of this material. The location of each Th atom corresponds to an atomic-scale dark dot at sites within the Uranium lattice, whose density is in good agreement with the known Th atom density and which are absent in undoped samples. In Figure 3*B-D* we show the $g(\varepsilon)$ measured under three different conditions. Fig. 3*B* shows the spatial average of $g(\varepsilon)$ away from Th sites measured above the temperature where the heavy fermion bands are formed (7,24,25). Fig. 3*C* shows the same spatial average $g(\varepsilon)$ measured deep in the heavy fermion state at 1.9 Kelvin but away from Th sites. We see clearly the changes in $g(\varepsilon)$ which are produced by the appearance of the coherent heavy fermion band structure, and the resulting practical definition of $\Delta_h$ as the difference between the energies where $N(\varepsilon)$ departs from its high-temperature values when heavy fermion bands are absent (green arrows). Fig. 3*D* shows the typical $g(\varepsilon)$ observed at each Th atom as measured at 1.9K; it always exhibits a quite sharp (δE~1meV) peak in $N(\varepsilon \sim 0)$. This is the long-sought electronic impurity bound-state at a Kondo-hole (19-23).

**Hybridization Gapmap.** The local hybridization strength in **r**-space is determined by measuring the range of energies between $\Delta_h^-(\mathbf{r})$ and $\Delta_h^+(\mathbf{r})$ for which $g(\mathbf{r}, \varepsilon)$ is distinct from the featureless $g(\mathbf{r}, \varepsilon)$ observed before the hybridization-split heavy bands appear (schematically in Fig. 1*B,C*; green arrows in Fig. 3*C* and Fig. 4*C*). To visualize the spatial arrangements of hybridization we therefore introduce the concept of a 'hybridization gapmap' $\Delta_h(\mathbf{r}) = \Delta_h^+(\mathbf{r}) - \Delta_h^-(\mathbf{r})$ for heavy fermion compounds (by analogy to the 'superconducting gapmap' technique (31)). Recall that this corresponds closely to the theoretical definition (4) of $\Delta_h(\mathbf{r})$ i.e. the difference between the energies of the heavy-fermion bands $\varepsilon^-(\mathbf{k})$ and $\varepsilon^+(\mathbf{k})$ measured at $\mathbf{k}^c(\varepsilon^f)$ (Fig. 1B) A typical result for $\Delta_h(\mathbf{r})$ in a 40x40 nm square FOV is shown in Fig. 3E (topograph in SI text). The hybridization strength is highly heterogeneous at the nanoscale ranging in energy over ~5mV. The inset shows how the hybridization oscillations are actually centered on the Th atoms. The unexpectedly intense and widespread hybridization disorder (Fig. 3E) then stems from the long-range nature of these hybridization oscillations in combination with the randomness of the Th sites upon which they are centered. Although hybridization disorder has long been anticipated for randomly Kondo-hole doped heavy fermion systems (12-16,19-23) it is compelling to observe it directly for the first time and understand its origin and intensity in terms of long-range hybridization oscillations.

## Discussion and Conclusions

Fourier analysis of hybridization gapmaps in Th-doped URu$_2$Si$_2$ (e.g. Fig. 3*E*) reveals that the apparently complex fluctuations in $\Delta_h(\mathbf{r})$ actually contain a quite simple structure. The Fourier transform, $\Delta_h(\mathbf{q})$, of $\Delta_h(\mathbf{r})$ from Fig. 3*E* is shown in Fig. 4*A* (see SI text). Here we see that the hybridization strength is modulating in space with a characteristic wave vector $\mathbf{Q}^*$ as identified by the arrow. This observation can be compared with the theoretical predictions from Fig. 2*C* (22) that hybridization modulations should exist at each Kondo-hole. In Fig. 4*B,C* we show the heavy fermion band structure of this material measured using quasiparticle interference imaging (7), along with associated changes in $N(\varepsilon)$. This simple **k**-space structure is in excellent agreement with basic heavy fermion theory (3-5). Figure 4*C*, which uses the same vertical axis as Fig. 4*B*, shows the energy range $\Delta_h$ in which the changes in the $g(\varepsilon)$ due to hybridization are observed. This comparison is highly consistent with theoretical expectations for the relationship between **k**-space and **r**-space electronic structure of a heavy fermion compound (Fig. 1*B,C*). The inset to Fig. 4*A* is the power spectral density measured along a typical radius in Fig. 4*A* showing that the oscillations of hybridization strength occur at the wavevector $|\mathbf{Q}^*|$ = 0.3±0.01 ($2\pi/a_0$) in good agreement with the **r**-space hybridization periodicity observed surrounding the Th sites (Fig. 3*E* inset). The oscillation wavevector is identified by the vertical dashed line in Fig. 4*B* thereby revealing that the hybridization oscillations occur at $\mathbf{Q} \cong 2\mathbf{k}_F^c$ (arrow) very close to where the light conduction band avoids its crossing with $\varepsilon^f$ and quite inconsistent with $\mathbf{Q} \cong 2\mathbf{k}_F^{HF}$ (arrow).

Long-standing theoretical predictions (19-23) for the electronic structure of a Kondo-hole including the suppression of hybridization near the substitution-atom sites (Fig. 3*D*), the disordered hybridization fluctuations generated by random Kondo-hole doping (Fig. 3*E*), that an impurity bound-state appears within the hybridization gap (Fig. 3*D*) and that hybridization oscillations exist and exhibit $\mathbf{Q} \cong 2\mathbf{k}_F^c$ (Fig. 4*A*) are borne out directly by these experiments. This agreement between theory (19-23), deductions from macroscopic experiments (12-18), and our observations at spinless Th atoms makes it highly implausible that these effects stem from another mechanism and, conversely, provides growing confidence in the ability to predict theoretically and to detect experimentally the atomic scale electronic structure and, perhaps more importantly, the consequent hybridization disorder generated by Kondo-holes. Future work will seek to establish any specific effects of the hidden order in $URu_2Si_2$ on the Kondo hole electronic structure and the universal characteristics of Kondo holes in a variety of other heavy fermion systems. Nevertheless, the advances we report here will be consequential for studies of nanoscale hybridization heterogeneity, local perturbations to heavy-fermion formation, Kondo-hole bound states, and suppression of coherence and superconductivity in Kondo-hole doped heavy fermion systems. The combination of SI-STM techniques introduced here and in Ref. 7 also provide a powerful approach for study of the **r**-space and **k**-space electronic structure of heavy fermion systems

**ACKNOWLEDGEMENTS** We acknowledge and thank P. Coleman, P. Chandra, T. Durakievich, J. Figgins, Z. Fisk, M. Graf, K. Haule, E.-A. Kim, G. Kotliar, D.-H. Lee, D. Morr, K. M. Shen, F. Steglich, Z. Tešanović, J. Thompson, M. Vojta and J.X. Zhu, for helpful discussions and communications. These studies at Brookhaven National Laboratory and Cornell University were supported by the U.S. Department of Energy, Office of Basic Energy Sciences. Research at McMaster was supported by the National Science and Engineering Research Council of Canada and the Canadian Institute for Advanced Research. Research at Los Alamos was supported by the U.S. Department of Energy Office of Basic Energy Sciences, Materials Sciences Division, and in part by the Center for Integrated Nanotechnology, a U.S. Department of Energy Office of Basic Energy Sciences user facility, under contract DE-AC52-06NA25396. IAF acknowledges support from Fundação para a Ciência e a Tecnologia, Portugal under fellowship number SFRH/BD/60952/2009.

**Figure Legend**

**Fig. 1.** Schematic of the **r**-space and **k**-space electronic structure of a simple heavy fermion system

(A) Schematic of a Kondo Lattice. The atoms/spins shown in blue represent the magnetic atoms. The conduction electrons which screen these moments at low temperature are indicated schematically in red.

(B) Schematic of a simple heavy fermion band structure in **k**-space. Here the (light) conduction electron band $\varepsilon^c(\mathbf{k})$ is hole-like and indicated by the steeply sloping dashed line. The two heavy fermion bands are indicated by solid lines, with the energy range of their avoided crossing, the hybridization gap $\Delta_h$, as shown.

(C) Schematic of the density of states $N(\varepsilon)$ simulated at finite temperature for a heavy fermion system as shown in b. We see clearly how $\Delta_h$ can be determined in **r**-space by identifying the energy range between $\Delta_h(\mathbf{r}) = \Delta_h^+(\mathbf{r}) - \Delta_h^-(\mathbf{r})$ within which the $N(\varepsilon)$ is perturbed from its structure before the heavy fermion bands appeared.

**Fig. 2.** Theoretical analysis of hybridization oscillations surrounding a Kondo hole

(A) Schematic of a single Kondo resonance. The magnetic atom is shown in blue while the conduction electrons which screen its moment at low temperature are indicated schematically in red. The non-magnetic atoms in the host metal are green.

(B) Schematic of a single Kondo-hole. The magnetic atom/spin is shown in blue while the conduction electrons which screen these moments at low temperature are indicated schematically in red. The non-magnetic atom at the Kondo-hole site is shown in green.

(C) Theoretical analysis of variations in the intensity of the hybridization strength and thus $\Delta_h$ surrounding a Kondo-hole (reproduced with permission from ref. 22). Here the wavevector of strongest perturbation $\mathbf{Q}^*$ is predicted to occur at $\mathbf{Q}^* \cong 2\mathbf{k}_F^c$.

(D) The conduction (small circular) and heavy fermion bands (large) used in the model of ref. 22 (reproduced with permission). It is obvious by comparison that the hybridization oscillations in $\Delta_h$ surrounding the Kondo-hole occur at the wavevector $\mathbf{Q}^* \cong 2\mathbf{k}_F^c$ and are highly distinct from the Friedel oscillation of at the heavy fermion Fermi wavevectors with wavevector $\mathbf{q} \cong 2\mathbf{k}_F^{HF}$.

**Fig. 3.** Visualization of the effect of Kondo holes on the r-space hybridization strength

(A) Topographic image of the Uranium-terminated (7) surface of URu$_2$Si$_2$. The locations of the Th atoms substituted on the U sites appear as dark dots whose density is in agreement with the 1% substitution level. The image has been Fourier filtered to clearly visualize the position of the Th atoms on the U lattice.

(B) The tunneling spectrum $g(\varepsilon)$ far from the Th sites when measured at T~19K well above the temperature where the heavy fermion bands are observed to form.

(C) The tunneling spectrum $g(\varepsilon)$ measured at T~2K after the heavy fermion bands have become established (7) and far from the Th sites. The magnitude of the hybridization gap $\Delta_h$ can be determined by identification of the energy range where $g(\varepsilon)$ has been perturbed due to the HF bands structure as show by vertical green arrows (see Fig. 1$B,C$).

(D) The tunneling spectrum $g(\varepsilon)$ measured at T~2K at representative Th atom sites. Here the changes in hybridization can be determined from the energy range where $g(\varepsilon)$ is perturbed. The new local electronic state bound within the hybridization gap at each Th-atom Kondo hole site is seen as a sharp peak in $g(\varepsilon)$ near $\varepsilon$~0.

(E) The heavy-fermion 'hybridization gapmap' shows the spatial variations in hybridization strength as indicated by the color scale shown in the inset. The image is a direct visualization of nanoscale spatial variations of the hybridization energy $\Delta_h(\mathbf{r})$ in a Kondo-hole doped heavy fermion compound. The inset shows directly how the hybridization oscillations are actually centered on the Th sites; it is the average of the lower hybridization gap edge $\Delta^-(\mathbf{r})$ (because this exhibits the highest contrast) over all locations centered on a Thorium atom.

**Fig. 4.** Structure of hybridization oscillations in $\mathbf{q}$-space

(A) The Fourier transform of data in Fig. 3$E$ revealing that the perturbations to hybridization in a random distribution of Kondo-holes are not completely random because of the hybridization oscillations surrounding each Th atom. The central peak of the Fourier transform which is generated by the randomness of Th dopant disorder has been removed for clarity. An example of the measured wavevector of the hybridization oscillations $\mathbf{Q}^*$ is shown by the red arrow. The four dots in the corner of the image occur at values of $\mathbf{q} =(0,2\pi/a_0);(2\pi/a_0,0)$ where $a_0$ is the unit cell dimension. The inset shows the quantitative measurement of $\mathbf{Q}^*$.

(B) The measured characteristics of the light conduction band $\mathbf{k}^c(\varepsilon)$ observed above the temperature of heavy-fermion formation, with its Fermi wavevector $\mathbf{k}_F^c(\varepsilon)$ are shown in red. In blue, the characteristics, measured using heavy quasiparticle interference (7), of the two heavy bands which are formed from the light conduction band by the avoided crossing. Here the heavy fermion Fermi wavevector is $\mathbf{k}_F^{HF}(\varepsilon)$. The theoretical definition (4) of $\Delta_h$ for heavy fermion bands $\mathbf{k}_\pm^c(\varepsilon)$ is shown by the horizontal dotted lines. The measured wavevector of the hybridization oscillations $\mathbf{Q}^*$ is shown by a vertical dashed line.

(C) The associated changes in conductance $\delta g(\varepsilon)$ measured by subtracting the $g(\varepsilon)$ at low temperature from that at high temperature. The plot shows how the evaluation in $\mathbf{r}$-space of $\Delta_h$ is carried out using the difference in energy between the energies $\Delta^+$ and $\Delta^-$ where $g(\varepsilon)$ departs from its value before heavy fermion formation (green arrows). (The plot also shows that $\Delta_h$ measured independently in this fashion is in very good agreement with that measured in $\mathbf{k}$-space).

Figure 1

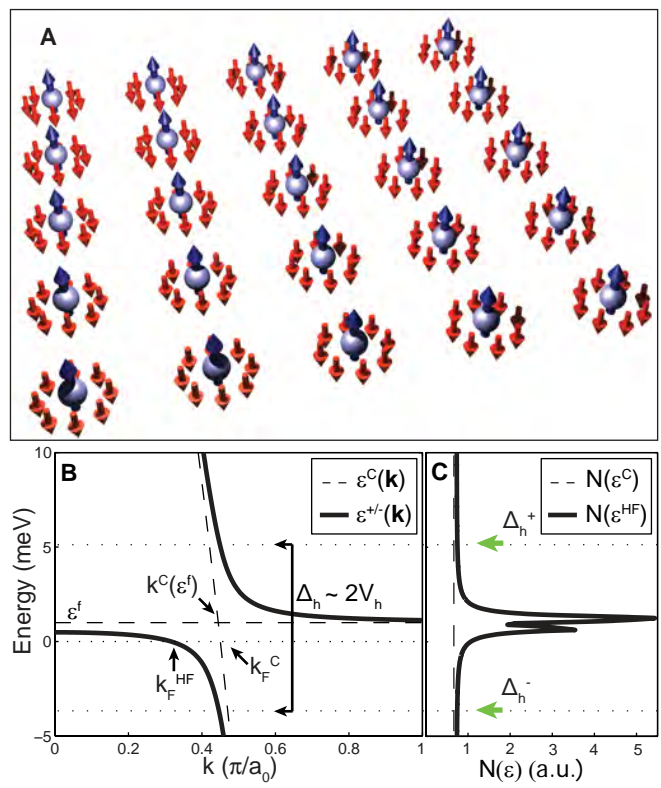

Figure 2

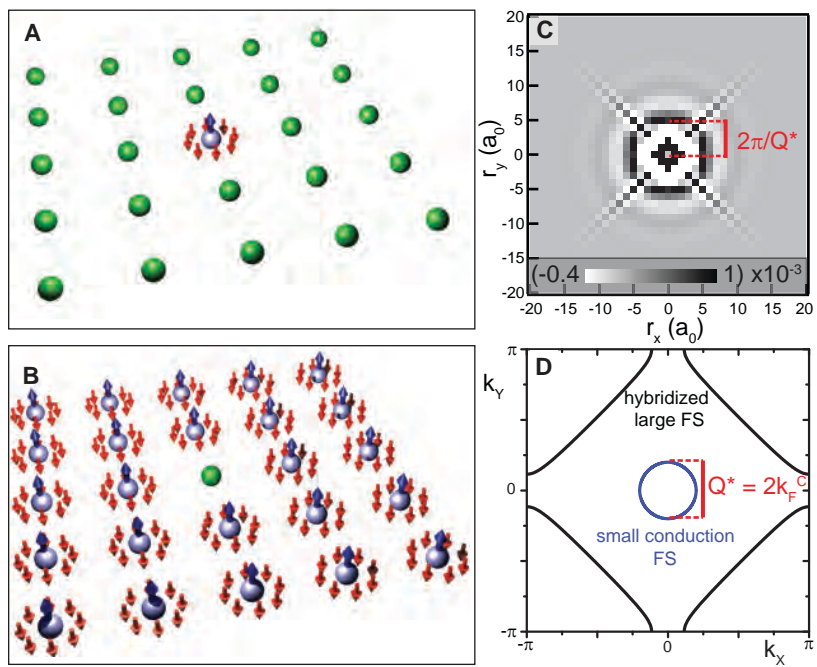

Figure 3

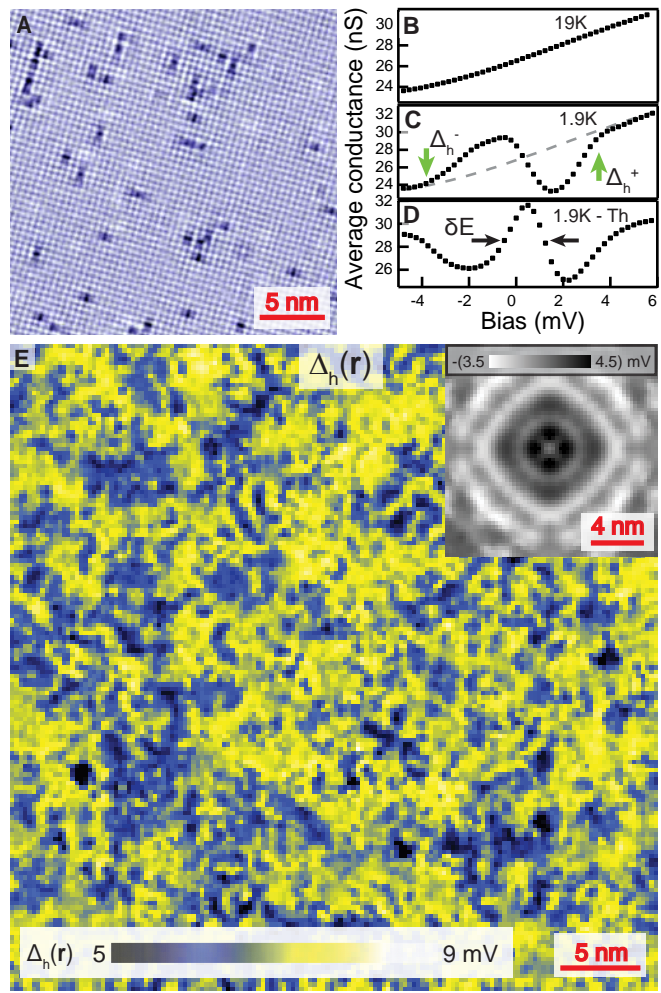

Figure 4

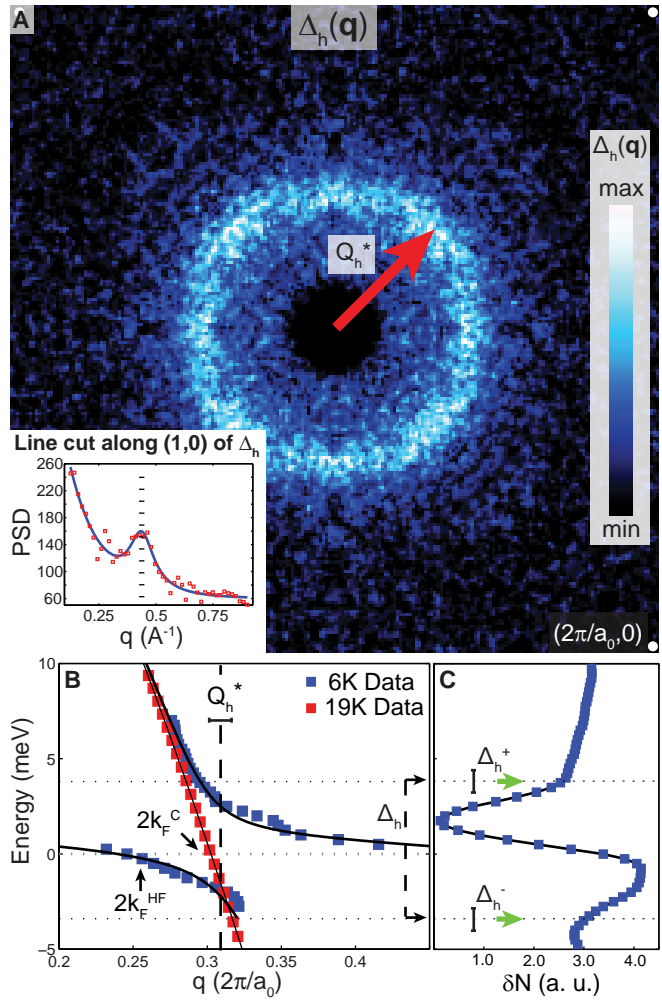

Supporting Information for

# How Kondo Holes Create Intense Nanoscale Heavy-Fermion Hybridization Disorder

M.H. Hamidian, A.R. Schmidt, I. Firmo, M.P. Allan, P. Bradley, J.D Garrett, T.J. Williams, G.M. Luke, Y. Dubi, A.V. Balatsky, and J.C. Davis

### (I) Simultaneous Topography of Data in Fig. 3*E*

In Fig. S1 we show the simultaneous topography obtained for the data present in Fig. 3*E*.

### (II) Hybridization Gap Algorithm

The hybridization gap in $URu_2Si_2$ is derived from STM spectroscopic measurements mapping the local value of the conductance in a 40nmx40nm field of FOV discretized across 200x200 pixels. As variations in the scale associated with the hybridization gap are small, high energy resolution is required, with the current data having an energy spacing of 250μV.

The edges of the hybridization gap, $\Delta_h^+$ and $\Delta_h^-$, are determined by finding the energies for which the low temperature conductance curve begins to deviate from the high temperature data, where no heavy fermion bands are observed, as discussed in the main text and Figs. 3*B* and *C*. It is found that the deviation points actually correspond very closely to inflection points on the density of states (DOS) curve. Thus, the algorithm to extract the energy position of the gap edges for the whole set of spectra (40 000 in total) involves fitting the interval around $\Delta_h^+$ and $\Delta_h^-$ with a low order polynomial and then determining the inflection points. This procedure generates the maps $\Delta_h^-(\mathbf{r})$ and $\Delta_h^+(\mathbf{r})$ for which the dominant modulation wavevectors can be seen by taking the PSD density as presented in Fig. S2. To increase the signal-to-noise ratio (S/N), the PSD has been symmetrised. Furthermore, the center has been removed to enhance the color contrast at the shorter wavelengths. As demonstrated in the power spectral density (PSD) plots of Fig. S2, each gap edge also modulates with Q* = 0.3 (2π/$a_0$) but with the dominant modulation wave vectors rotated by 45 degrees from one another. It is also found that $\Delta_h^-$ modulates more strongly than $\Delta_h^+$ by 30% and thus the contrast in the raw Fourier transform image of $\Delta_h(\mathbf{q})$ is dominated by weight corresponding to $\Delta_h^-$. To make the full ring of wavevectors have comparable contrast in the PSD, $\Delta_h(\mathbf{q})$ of Fig. 4*A*, the right edge PSD amplitude was augmented by 10%. The visually modified PSD for the full gap was then obtained by adding the images of the enhanced right edge PSD in quadrature with that of the left edge PDS, yielding the image in Fig. 4*A*.

The image modifications were only made to more clearly visualize the full ring of wavevectors which contribute to the hybridization oscillations. Quantitative analysis on the unmodified image of $\Delta_h(\mathbf{q})$, clearly show that there is a high S/N for determining Q* without the need for any alterations. Fig. S3, plots linecuts taken from the unprocessed PSD, $\Delta_h(\mathbf{q})$, along the (1,0) and (1,1) directions (red x marks) with enough S/N to pick out dominant wavevector. Quantitatively, the cuts were fit (blue curve) using an exponential background with a Lorentzian peak with the position of the Lorentzian determining the Q* value.

**Figure Legend**

**Fig. S1.** Simultaneous topography of the 40nmx40nm FOV for the data presented in Fig. 3*E*. Dark spots correspond to surface Th atoms. White spots correspond to subsurface impurities.

**Fig. S2.** Power spectral density of real space hybridization gap edge modulations.

PSD of (A) $\Delta_h^-(\mathbf{r})$ and (B) $\Delta_h^+(\mathbf{r})$. The dominant modulation vectors for both edges have the same magnitude, Q*, but their most intense direction are rotated 45 degrees from one another. The centers have been removed to enhance contrast.

**Fig. S3.** Linecuts along PSD images to extract Q*.

Linecuts along the (A) (1,0) and (B) (1,1) direction of the unmodified PSD, $\Delta_h(\mathbf{q})$. The modulation along (1,0) is stronger but both graphs clearly show the peak Q* originating from the hybridization oscillations.

**Figure S1**

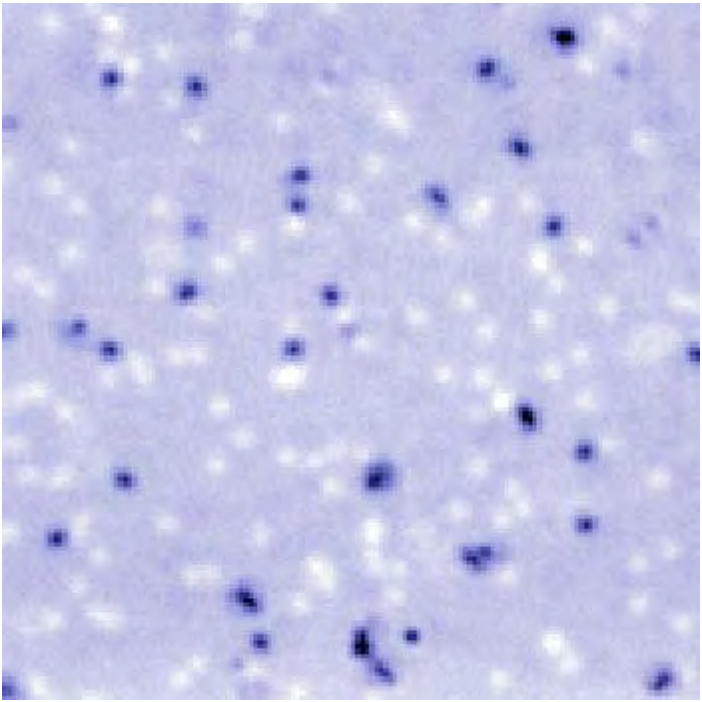

**Figure S2**

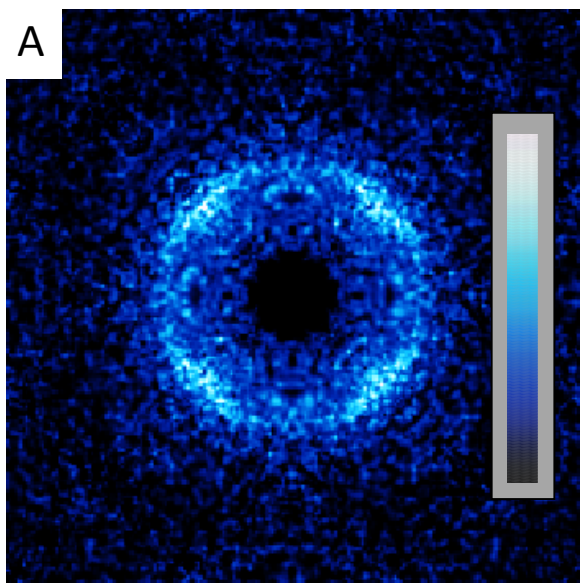 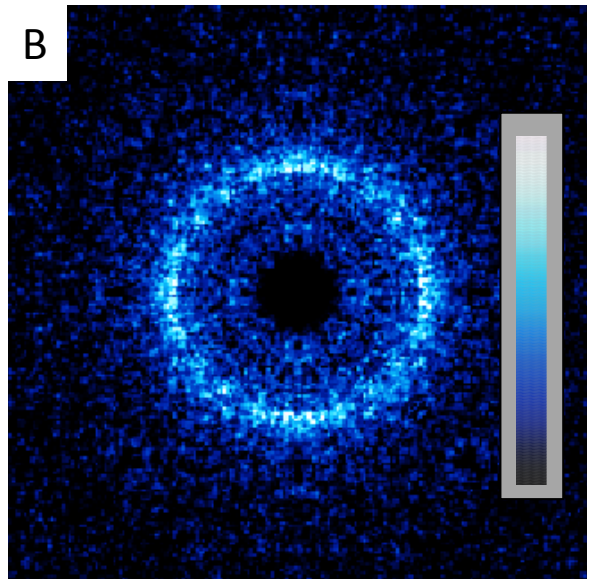

**Figure S3**

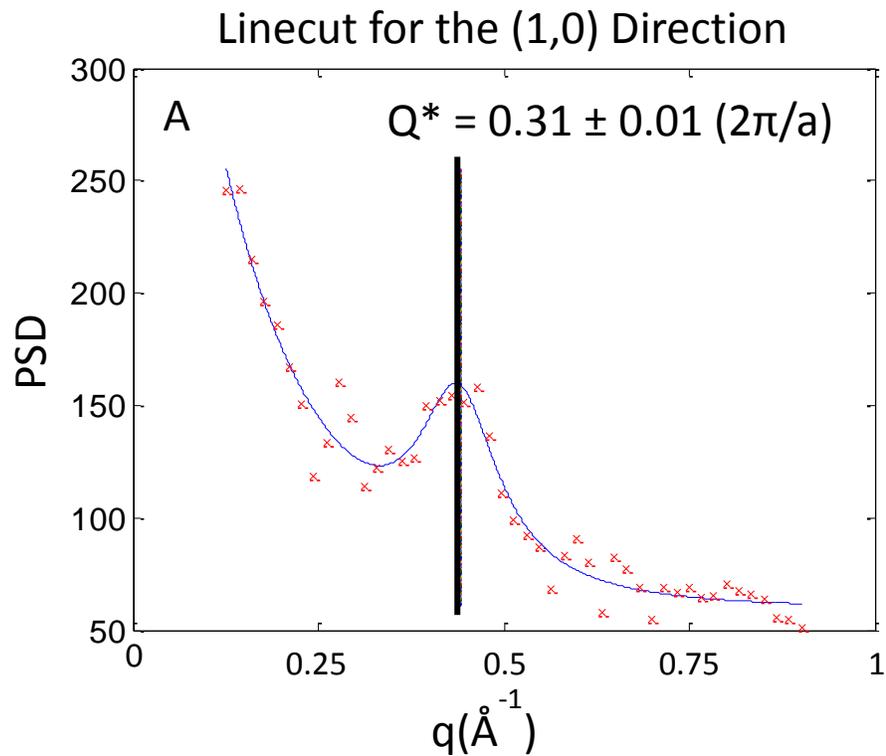

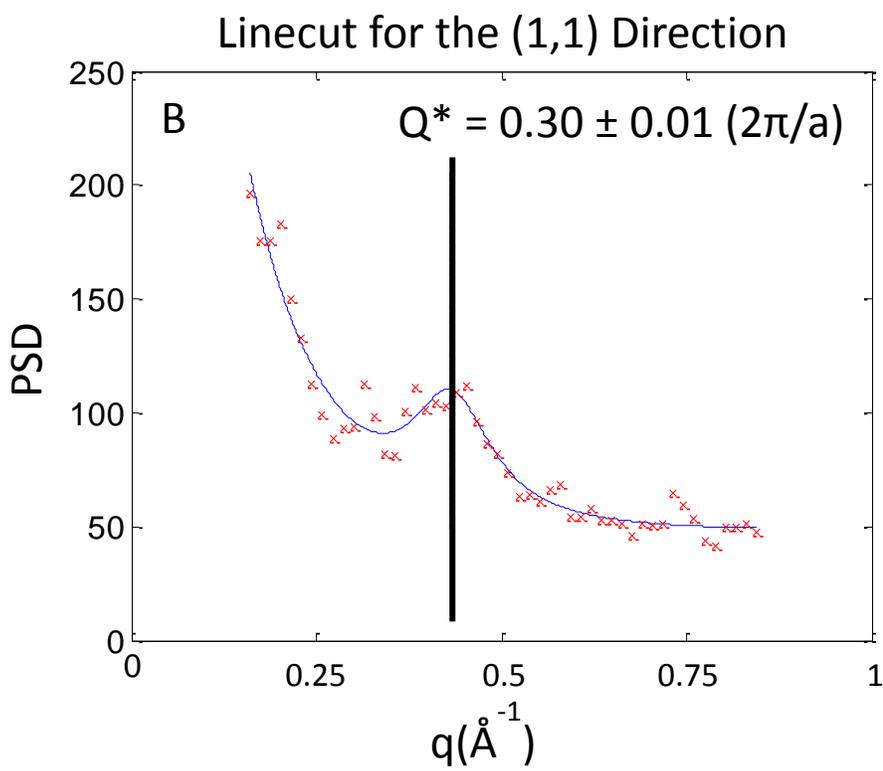